\documentclass[runningheads]{llncs}
\usepackage[T1]{fontenc}
\usepackage{graphicx}
\usepackage{graphicx}
\usepackage{xcolor}
\usepackage{comment}
\usepackage[utf8]{inputenc}
\usepackage{multirow}
\usepackage{arydshln}
\usepackage{subfigure}
\usepackage{float}
\usepackage{array}
\usepackage{tabularx}
\usepackage{graphicx}
\usepackage{url}
\usepackage{adjustbox}
\usepackage{xspace}
\usepackage{xcolor}
\usepackage{comment}
\usepackage{caption} 
\usepackage{adjustbox} 
\usepackage{booktabs}
\usepackage{hyperref}
\PassOptionsToPackage{bookmarks=false}{hyperref}
\usepackage{nicematrix}
\restylefloat{table}
\usepackage{latexsym}
\usepackage{amsmath, amssymb}
\usepackage{type1cm}        
\usepackage{comment}
\usepackage{makeidx}         
\usepackage{multicol}        
\usepackage[bottom]{footmisc}
\usepackage{newtxtext}       %
\usepackage[varvw]{newtxmath} 
\usepackage{type1cm}        
\usepackage{makeidx}         
\usepackage{graphicx}        
\usepackage{multicol}        
\usepackage[bottom]{footmisc}
\usepackage{newtxtext}       %
\usepackage[varvw]{newtxmath}       
\usepackage{color,array}
\usepackage{comment}
\usepackage{subcaption}
\usepackage{soul,xcolor} 
\usepackage{mdframed} 

\newcommand{\orcid}[1]{\href{https://orcid.org/#1}{\includegraphics[width=8pt]{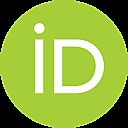}}}
\definecolor{highlightgreen}{rgb}{0.85, 1.0, 0.85} 
\definecolor{highlightorange}{rgb}{1.0, 0.9, 0.6} 
\definecolor{highlightred}{rgb}{1.0, 0.8, 0.8}  
\definecolor{highlightblue}{rgb}{0.85, 0.9, 1.0}  
\definecolor{highlightyellow}{rgb}{1.0, 0.95, 0.5} 

\definecolor{backgG}{RGB}{255, 255, 153}
\definecolor{tagtxtG}{RGB}{102, 102, 0}
\definecolor{backgPc}{RGB}{179, 255, 179}
\definecolor{tagtxtPc}{RGB}{0, 102, 0}
\definecolor{backgPw}{RGB}{255, 179, 179}
\definecolor{backgPw}{rgb}{0.0, 1.0, 1.0}
\definecolor{tagtxtPw}{RGB}{0.0, 1.0, 1.0}
\definecolor{backgPo}{rgb}{0.76, 1, 1}
\definecolor{tagtxtPo}{RGB}{102, 0, 0}
\definecolor{backgPm}{rgb}{0.98, 0.81, 0.69}
\definecolor{tagtxtPm}{RGB}{0,1,1}

\makeindex 

\begin{document}
\title{Overview of the SciHigh Track at FIRE 2025: Research Highlight Generation from Scientific Papers}
%
%
\author{Tohida Rehman\inst{1} \orcid{0000-0002-3578-1316} \and
Debarshi Kumar Sanyal\inst{2} \orcid{0000-0001-8723-5002} \and
Samiran Chattopadhyay\inst{1, 3} \orcid{0000-0002-8929-9605}}

\authorrunning{T. Rehman et al.}
\institute{Jadavpur University, Kolkata-700106, India.\\   \email{tohidarehman.it@jadavpuruniversity.in}\\
\and
Indian Association for the Cultivation of Science, Kolkata-700032, India.\\
\email{debarshi.sanyal@iacs.res.in}\\
\and 
Techno India University, Kolkata-700091, India. 
\email{samirancju@gmail.com}
}
%

\maketitle
\begin{abstract}\unskip

`SciHigh: Research Highlight Generation from Scientific Papers' focuses on the task of automatically generating concise, informative, and meaningful bullet-point highlights directly from scientific abstracts. The goal of this task is to evaluate how effectively computational models can generate highlights that capture the key contributions, findings, and novelty of a paper in a concise form. Highlights help readers grasp essential ideas quickly and are often easier to read and understand than longer paragraphs, especially on mobile devices. The track uses the MixSub dataset \cite{10172215}, which provides pairs of abstracts and corresponding author-written highlights.

In this inaugural edition of the track, 12 teams participated, exploring various approaches, including pre-trained language models, to generate highlights from this scientific dataset. All submissions were evaluated using established metrics such as ROUGE, METEOR, and BERTScore to measure both alignment with author-written highlights and overall informativeness. Teams were ranked based on ROUGE-L scores. The findings suggest that automatically generated highlights can reduce reading effort, accelerate literature reviews, and enhance metadata for digital libraries and academic search platforms. SciHigh provides a dedicated benchmark for advancing methods aimed at concise and accurate highlight generation from scientific writing.


\end{abstract}

\section{Introduction}
An abstract is a short, coherent, and concise summary of a scientific paper that reflects its key contributions. Similarly, research highlights provide a compact, bullet-point version of the paper’s main findings, complementing the abstract. Text summarization aims to capture the essential content of one or more documents while reducing the reader’s effort in identifying key information. Automatic summarization methods generally fall into two categories: extractive and abstractive approaches \cite{luhn1958automatic,9623462}.

The SciHigh track at FIRE 2025 focuses on the task of automatic research highlight generation. Within this track, participants were invited to generate concise, informative bullet-point highlights directly from scientific abstracts. Currently, many publishers require authors to provide bullet-point highlights alongside abstracts. The number of scientific publications roughly doubles every 9 years \cite{van2014global}, resulting in a huge volume of papers across fields and subfields. For researchers, particularly novices, keeping up with all new research is time-consuming and challenging. Therefore, automatic systems capable of generating human-like highlights from scientific papers are increasingly necessary.

In its first edition, 12 teams participated in this task, exploring various approaches to generate highlights from the provided dataset. The main goals of this track are to:
\begin{enumerate}
    \item Help readers quickly identify the main contributions of a scientific paper. Highlights are easier to read and understand than longer descriptive paragraphs, particularly on mobile or hand-held devices.
    \item Reduce the effort and time required to locate essential information within research articles.
    \item Improve the quality of metadata used by academic search engines and digital libraries.
    \item Provide a basis for evaluating the performance of different summarization methods on scientific texts.
\end{enumerate}

\section{Related Work}
Natural language generation has advanced rapidly with the rise of deep learning, particularly transformer-based models \cite{vaswani2017attention}, which play a key role in tasks such as translation, summarization, and question answering. Early work on extractive summarization was introduced by Luhn \cite{luhn1958automatic}, who ranked sentences based on word-frequency while ignoring common terms. Abstractive summarization has since evolved from attention-based seq2seq and convolutional models to pointer-generator networks with coverage mechanisms and richer embeddings, as well as approaches using fuzzy rules and genetic algorithms across multiple datasets \cite{Sutskever-2014-sequence,See2017GetTT}.

Scientific paper summarization can be broadly categorized into generating abstracts from the paper itself, producing summaries based on citations \cite{altmami2020automatic}, and the more recent task of generating bullet-point research highlights. Early approaches primarily relied on extractive methods. For instance, Kupiec et al. \cite{kupiec1995trainable} used a small dataset of 188 paper-summary pairs and ranked sentences based on selected features, while Contractor et al. \cite{contractor-etal-2012-using} employed argumentative zones for extractive summarization. Kinugawa and Tsuruoka \cite{kinugawa2018hierarchical} introduced a two-level hierarchical encoder-decoder model for scientific papers. With the increasing demand for research highlights alongside abstracts, highlight generation has become a distinct task within scientific text summarization. Collins et al. \cite{collins2017supervised} proposed a supervised extractive model to identify highlight sentences and contributed CSPubSum, a benchmark dataset of computer science publications. Cagliero et al. \cite{cagliero2020extracting} developed a gradient boosting-based method to extract highlights, evaluating it on CSPubSum as well as two domain-specific datasets, AIPubSumm and BioPubSumm, curated from ScienceDirect using AI and biomedical-focused keyword queries. Rehman et al. \cite{rehman2021automatic} introduced an abstractive method for generating highlights using a pointer-generator model with GloVe embeddings, later enhanced with named entity recognition \cite{rehman-etal-2022-named} and contextual ELMo embeddings \cite{rehman2023research,peters-etal-2018-deep}. More recently, they combined SciBERT embeddings with a pointer-generator network using a coverage mechanism and proposed the MixSub multi-disciplinary dataset, containing research articles from multiple domains \cite{10172215}. 

This track will evaluate the performance of approaches for generating highlights from the MixSub dataset, providing a benchmark for comparing methods across domains.

\section{Dataset}
\label{datasets}
For the highlight generation task, we use the \textbf{MixSub} dataset \cite{10172215}, which contains 19,785 research articles from multiple domains published in 2020 on ScienceDirect\footnote{\url{https://www.sciencedirect.com/}}. The dataset includes articles from the Biological Sciences, Chemistry, Energy, Management, Nursing, Physics, and Social Science domains, drawn from various journals published in that year. Each article includes an \textit{abstract} and corresponding author-written \textit{highlights}. The dataset is split into training, validation, and test sets in an 80:10:10 ratio. For the SciHigh track at FIRE 2025, we sampled 10,000 instances for training out of 15,960 available instances, 1,985 for validation, and 1,840 for testing. Table \ref{Table:sample_MixSub} shows an example entry from MixSub.
\begin{table*}[!h]
\centering
\caption{\small Example of an \textit{abstract} and its corresponding \textit{author-written highlights} from the MixSub dataset (\url{https://www.sciencedirect.com/science/article/abs/pii/S0001457519303616}). Colored boxes indicate the relationship between segments of the abstract and the highlights. In some cases, highlights directly reflect portions of sentences in the abstract, while in others, a highlight may combine information from multiple sentences and rephrase it with different wording. This illustrates the challenge of mapping abstracts to highlights for automatic generation tasks.}
	
\label{Table:sample_MixSub} 
\resizebox{\linewidth}{!}{
{\begin{tabular}{ |p{12.5 cm}|} \hline
{\bf Abstract:} ``We propose \colorbox{blue!10}{a novel network screening method for hotspot identification} based on the optimization framework to maximize the total summation of a selected safety measure for all hotspots considering a resource constraint for conducting detailed engineering studies. The proposed method allows the length of each hotspot to be determined dynamically based on constraints the users impose. The calculation of \colorbox{pink!30}{the Dynamic Site Length} method is based on Dynamic Programming and it is shown to be effective to find the close to optimal solution with computationally feasible complexity. The screening method has been demonstrated using historical crash data from extended freeway routes in San Francisco California. Using the Empirical Bayesian estimate as a safety measure we compare the performance of the proposed DSL method with other conventional screening methods Sliding Window and Continuous Risk Profile in terms of their optimal objective value. Moreover their spatio \colorbox{yellow!20}{temporal consistency} is compared through the site and method consistency tests. \colorbox{cyan!20}{Findings show that DSL can outperform SW and CRP} in investigating more hotspots under the same amount of resources allocated to DES by pinpointing hotspot locations with greater accuracy and showing improved spatio temporal consistency.''\\\hline 	    
{\bf Author-written research highlights:} 
\begin{itemize}
    \item[$\blacktriangleright$] ``\colorbox{blue!10}{A novel network screening method for hotspot identification is proposed.}''
    \item[$\blacktriangleright$] ``\colorbox{pink!30}{The Dynamic Site Length} DSL method allows of a dynamic hotspot length.''
    \item[$\blacktriangleright$] ``A budget constraint for site investigation is considered.''
    \item[$\blacktriangleright$] ``\colorbox{cyan!20}{Three network screening methods are tested.}''     
    \item[$\blacktriangleright$] ``DSL has higher spatial \colorbox{yellow!20}{temporal consistency} than other existing screening methods.''
\end{itemize}\\\hline
\end{tabular} }

}
\end{table*}	

\section{Task Description}
The SciHigh track focuses on the task of automatically generating research highlights from scientific paper abstracts using the MixSub dataset \cite{10172215}. While abstracts summarize a paper, research highlights provide a more structured and concise representation of its key contributions. The goal of this task is to develop machine learning models capable of producing high-quality highlights that closely resemble those written by authors.

Participants can explore various approaches, including transformer-based models, retrieval-augmented methods, and fine-tuned neural networks. 

The task is intended to advance the efficiency and quality of automatic highlight generation, supporting researchers in quickly grasping key contributions and enhancing academic search and indexing platforms.

\section{Participation and Evaluation}

\begin{table*}[!htp]
\centering
\caption{Performance of participating teams in the SciHigh track based on ROUGE-L F1 scores. The best run for each team is reported.}
\label{Table:results}
\begin{tabular}{l c c c}
\toprule
\textbf{Group Name} & \textbf{Run Submission (Best Run)} & \textbf{ROUGE-L F1} & \textbf{Rank} \\
\midrule
Text\_highlights\_gen & run1 & 23.45 & 1 \\
AiNauts & run1 & 23.24 & 2 \\
SVNIT\_CSE & run1 & 23.02 & 3 \\
NLPFusion & run2 & 22.96 & 4 \\
The NLP Explorers & run2 & 22.94 & 5 \\
NIT\_PATNA\_2025 & run1 & 22.42 & 6 \\
MUCS & run1 & 22.08 & 7 \\
JU\_CSE\_PR\_KS & run1 & 22.06 & 8 \\
SCaLAR & run1 & 20.33 & 9 \\
Ayanika & run1 & 17.91 & 10 \\
\bottomrule
\end{tabular}
\end{table*}

Initially, 14 teams from various institutions, universities, and colleges registered for the SciHigh track. Out of these, 12 teams submitted their approaches along with trained models, each providing up to two runs in CSV format containing predicted highlights. The test set, consisting of 1,840 instances, was initially released with masked ground-truth highlights. After submission, the full test set with author-written highlights was released for evaluation.

All submissions were evaluated using ROUGE-1, ROUGE-2, and ROUGE-L metrics. To ensure consistency, we provided the same evaluation code to all teams for verification of their model performance. The final ranking of submissions was based on ROUGE-L F1 scores.  

Teams employed a variety of strategies, including hybrid approaches, extractive methods, and fine-tuning of pre-trained language models. Table \ref{Table:results} summarizes the results, showing the performance of all participating systems according to ROUGE-L F1 scores. 

The \textbf{Text\_highlights\_gen} team developed two models: \texttt{Pegasus} and \texttt{Pegasus} enhanced with \texttt{NER} features. Both models were fine-tuned for 10 epochs with a batch size of 2, using the Adam optimizer and a learning rate of $2\mathrm{e}{-}5$. Beam search with a width of 4 was adopted to improve sequence generation quality. Inputs were truncated to 512 tokens, and outputs were limited to 100 tokens. The fine-tuned Pegasus model achieved a ROUGE-L F1 score of 23.45\%, securing the 1st position in the SciHigh track of FIRE 2025 conference.

The \textbf{AiNauts} team employed two approaches for highlight generation. Method 1 combined extractive and abstractive steps, ranking sentences using TF–IDF, Sentence-BERT embeddings, cosine similarity, and MMR, followed by abstractive rewriting with a fine-tuned \texttt{Facebook/bart-large-cnn} model. Method 2 used a fine-tuned \texttt{DistilBERT-base-uncased} model for binary sentence classification, selecting sentences with probabilities above 0.5. Both models were trained for 3 epochs, with batch size 8 for Method 1 and batch size 16 for Method 2. Their best result came from Method 1, achieving a ROUGE-L F1 score of 23.24\% and ranking 2nd in the SciHigh track this year.

Team \textbf{SVNIT\_CSE} used ensemble of transformer based models for the task of highlight generation \texttt{facebook/bart-large-cnn}, \texttt{google/pegasus-pubmed}, two variants of T5 such as \texttt{t5-base}, \texttt{google/long-t5-tglobal-base}, and \texttt{allenai/led-base-16384}. BART, T5, and Long-T5 were fine-tuned with a batch size of 8, while LED and Pegasus used a batch size of 4. Maximum input lengths were set to 384 (BART), 512 (T5), 516 (Long-T5), 2048 (LED), and 516 (Pegasus), with outputs limited to 64 tokens. Their best performance came from the \texttt{bart-large-cnn} model, achieving a ROUGE-L F1 score of 23.02\% and placing the team 3rd among 12 participants in the SciHigh track.

Team  \textbf{NLPFusion} fine-tuned Pegasus model using Low-Rank Adaptation (LoRA) for abstractive summarization. They evaluated both \texttt{Pegasus-PubMed} and \texttt{Pegasus-ArXiv}, tokenizing inputs to 256 tokens and outputs to 64 tokens. Both variants were fine-tuned under identical settings for fair comparison. Their best result came from the Pegasus-PubMed + LoRA model, which achieved a ROUGE-L F1 score of 22.96\%, placing the team 4th among all the participants in this track.

Team  \textbf{The NLP Explorers} employed the \texttt{T5-base} and \texttt{BART-base} models, fine-tuning it for 5 epochs with a batch size of 8 and a learning rate of $2\mathrm{e}{-}5$. Input abstracts were truncated to a maximum length of 512 tokens, while generated highlights were limited to 100 tokens. Their T5-base fine-tuned model achieved its best performance, with a ROUGE-L F1 score of 22.94\%, ranking 5th in the SciHigh track.

Team \textbf{NIT\_PATNA\_2025} worked under two sub-teams but later merged their findings into a single submission. They experimented with \texttt{T5-small} and a \texttt{LongT5} models, both fine-tuned for 10 epochs using the same settings as that of the \textbf{Text\_highlights\_gen} team. Their best performance came from the \texttt{T5-small} model, which achieved a ROUGE-L F1 score of 22.42\%, placing them 6th among the track participants.

Team  \textbf{MUCS} fine-tuned a \texttt{T5-base} model for 2 epochs, with a maximum input length of 512 tokens, an output limit of 128 tokens, a learning rate of $3\mathrm{e}{-}4$, and a batch size of 4 for both training and evaluation. Their system achieved a ROUGE-L F1 score of 22.08\%, placing the team 7th in the SciHigh track.

Team  \textbf{JU\_CSE\_PR\_KS} experimented with two models: an XGBoost classifier and an XGBoost regressor. The classifier assigned binary labels to sentences based on their overlap with reference highlights, while the regressor generated graded similarity scores (e.g., 0.4, 0.9, 0.6) for finer ranking. The top-scoring sentences were selected as highlights. Their regressor model achieved a ROUGE-L F1 score of 22.06\%, placing the team 8th in the SciHigh track of FIRE~2025.

Team \textbf{SCaLAR} proposed an automated highlight-generation pipeline integrating entity extraction with SciBERT, keyword extraction with KeyBERT, sentence ranking via token budgeting, and supervised fine-tuning of LLaMA. They also explored a retrieval-augmented setup using BART, SciBART, and T5 with Facebook AI Similarity Search(FAISS) based similar-example retrieval to guide generation. Their best configuration (V6), combining trimmed abstracts, guided constraints, and reference-aligned filtering, achieved a ROUGE-L F1 score of 20.33\%, placing the team 9th in the track.

Team \textbf{Ayanika} fine-tuned pretrained language models, specifically \texttt{T5-small} for highlight generation. Their best system achieved a ROUGE-L F1 score of 17.91\%, placing the team 10th in the SciHigh track of FIRE 2025.

\noindent Remaining two teams did not submit their working notes for the track.

\section{Conclusion and Future Scope}
The first sub-task of the SciHigh track at FIRE 2025 aimed to generate research highlights from the MixSub dataset. Teams used a variety of methods, including extractive strategies, hybrid approaches, and fine-tuned pre-trained language models. The results show that current techniques can produce concise and informative highlights, but challenges remain, such as effectively combining information from multiple sentences and maintaining the original meaning of the text.

For future work, research can focus on generating highlights across different scientific domains, handling multiple languages, and incorporating external knowledge or retrieval-based methods to improve accuracy and coverage. The track establishes a benchmark for automatic highlight generation and contributes to making scientific literature more accessible to researchers and academic platforms.

\bibliographystyle{unsrt}
\bibliography{llm_ref}
\end{document}